# Biased Risk Parity with Fractal Model of Risk

Sergey Kamenshchikov, *PhD, IFCM Capital/Moscow Exchange*
Ilia Drozdov, *CFA, QB Capital*

**Abstract**

For the past two decades, investors have observed long-memory and highly correlated behavior of asset classes that doesn't fit into the framework of Modern Portfolio Theory. Custom correlation and standard deviation estimators consider normal distribution of returns and market efficiency hypothesis. It forced investors to search more universal instruments of tail risk protection. One of the possible solutions is a naive risk parity strategy, which avoids estimation of expected returns and correlations. The authors develop the idea further and propose a fractal distribution of returns as a core. This class of distributions is more general as it does not imply strict limitations on risk evolution. The proposed model allows for modifying a rule for volatility estimation, thus, enhancing its explanatory power. It turns out that the latter improves the performance metrics of an investment portfolio over the ten-year period. The fractal model of volatility plays a significant protective role during the periods of market abnormal drawdowns. Consequently, it may be useful for a wide range of asset managers which incorporate innovative risk models into globally allocated portfolios.

**Introduction**

The market crashes of 2000s have raised a question about applicability of Modern Portfolio Theory (MPT), as correlation and standard deviation of returns do not consider abnormal volatility [1]. In the long-term perspective, the infrequent large-scale declines can sufficiently affect the portfolio performance. In 2006-2016, the S&P500 drawdown of 36% has decreased the accumulated ten-year return two-fold and annual Sharpe ratio three-fold. Naive risk parity approach partially reduces the gap by avoiding calculation of expected returns and linear correlations. The global allocation strategy based on naive risk parity strategy has outperformed S&P500 index by 2% for the past forty years with annual Sharpe ratio increase of 63% [2]. However, the volatility estimation of global risk parity model is still based on normal distribution of returns that fails to explain such effects as collective behavior and long-term memory during market downturns [2]. It has been shown that fractal distributions provide more realistic approach for large-scale declines of asset values [3]. In this research, we introduce a fractal law of volatility evolution into the biased risk parity buy-and-hold portfolio. We compare the performance of proposed strategy with naive risk parity according to out-of-sample analysis of annual and half-year returns.

**Considerations for risk management**

Collective behavior and the fat-tailed distribution of returns have been in focus in financial research for a while. Mandelbrot proposed alpha-stable distributions because they admit the infinite volatility and large scale declines [4]. According to [3], the alpha stable distribution of returns is more general as it can produce the normal distribution but does not require finite variance of returns. Therefore, this model fits to efficient and inefficient regimes simultaneously. Central Limit Theorem supports the stability of this class of distributions [3]. Probability alpha stable distribution function of returns *r* is characterized by the following relation:

$$F(r\,|\,\alpha,\beta,t) = \frac{1}{\pi}\int_0^\infty \cos(h(r\,|\,\alpha,\beta,t))\exp(-t^\alpha)dt \qquad (1)$$

Here $\alpha, \beta, \sigma, \mu$ are static parameters of distribution, $t$ is time. The core of the integral is represented by the piecewise function $h$:

$$h(r \mid \alpha, \beta, t) = rt + \beta \tan\left(\frac{\pi \alpha}{2}\right)(t - t^\alpha), \alpha \neq 1 \quad (2)$$

$$h(r \mid \alpha, \beta, t) = \mu + \frac{2\beta\sigma}{\pi} \ln \sigma, \alpha = 1 \quad (3)$$

The degree of stability of this distribution is defined by $\alpha$ parameter that lies in the interval $0 < \alpha \leq 2$. Rescaling of this distribution with rescale factor $k$ corresponds to the following [4]:

$$F(r \mid \alpha, \beta, kt) = k^H F(r \mid \alpha, \beta, t) \quad (4)$$

Alpha parameter quantifies fat tails, thus, making it possible to consider them in volatility analysis. According to Mandelbrot [4], this key parameter is also directly related to Hurst factor $H$ of time series persistency: $\alpha = 1/H$. We should recall that positive memory (momentum) of returns is characterized by $H \in (0.5, 1]$, while negative memory (mean reversion) corresponds to $H \in [0, 0.5)$. The law for rescaling of volatility may be derived directly from (4):

$$E\left[(r(t_2) - r(t_1))^2\right] = |t_2 - t_1|^{2H} \quad (5)$$

Here the values of two consequent time moments $t_1, t_2 : t_2 > t_1$ are involved. In this research, the volatility is estimated as the unbiased standard deviation of returns. Accordingly, One can rescale one-period standard deviation $STD_0$ to derive an N-period standard deviation in the following way: $STD_N = STD_0 N^H$ [4]. MPT and market efficiency imply a random walk of returns and the short memory: $H = 0.5$ for any asset class. As a result, this rescaling factor does not play a role in the definition of investment weights up to a constant. In contrast, this research implies variable rescaling depending on asset type and time series historical properties. Here we apply an approach of small data basis, suggested in [5] for the stable definition of Hurst factor $H$.

**Investment strategy**

In this research, we suggest global diversification among four asset classes that are typically considered in Global Asset Allocation models. However, we admit the possibility of wider selections. We use adjusted daily close prices of several low-cost exchange traded funds (ETFs). All corporate actions such as dividends, splits etc. have been taken into account.

| Asset class | Index | Tradable ETFs | Expense ratio, % |
|---|---|---|---|
| Equities | S&P 500 | SPY | 0.09 |
| Treasuries | U.S. 20+ Year Treasury Bonds | TLT | 0.15 |
| Real Estate | U.S. Real Estate Index | IYR | 0.43 |
| Gold | Gold Bullion | GLD | 0.4 |

Table 1. Investment blocks

The applied historical data is limited by the inception date of ETFs, thus, the period of 2005-2016 is the core of the algorithm justification. The optimization period $N$ and the holding period are the same – we consider half-year and annual horizons independently. Naive risk parity approach implies that an equal amount of volatility is designated to each class. However, in this research, we propose trend-following bias and consider investments with positive expected return for the long-only portfolio. Below we compare fractal and standard algorithms of portfolio optimization.

| # | Standard biased | Fractal biased |
|---|---|---|
| 1 | Historical returns of asset $i$ are calculated as the difference of adjusted price logarithms $r_i = 100\% \cdot \Delta \log(p_i)$ ||
| 2 | Hurst factor $H_i = 0.5$ for each asset class | Hurst factor $H_i$ for each asset is defined based on daily returns time series |
| 3 | Expected daily return $\mu_i = mean(r_i)$ is defined as simple average of returns. The asset class weight is initially switched to zero if $\mu_i \leq 0$ ||
| 4 | Daily standard deviations of returns $STD_0^i$ are calculated ||
| 5 | Expected asset volatility is rescaled to the horizon $N$ of the portfolio: $STD_N^i = STD_0^i \sqrt{N}$ | Expected asset volatility is rescaled to the horizon $N$ of the portfolio: $STD_N^i = STD_0^i N^{H_i}$ |
| 6 | Non-zero investment weights are calculated based on rescaled volatilities $w_i = 1/STD_N^i$, the sum of weights is normalized: $w_i \to w_i / \sum w_i$ ||

Table 2. Investment scheme

In this research, non leveraged investment scheme is applied. However we admit leveraging in order to normalize expected portfolio volatility to the recommended level. Some researchers of risk parity models use benchmark (SPY) expected volatility as a reference point [6] or include independent risk aversion recommendations.

**Practical application**

We compare the simulated results of the three approaches – fractal biased, standard biased, and naive risk parity – in the Tables 3 and 4. Both algorithms are compared by several metrics: Sharpe ratio, Treynor ratio, average annual return, capital protection, standard deviation (STD) and beta. We find it reasonable to take one of the most popular proxies for the US stock market, SPY, as a benchmark. All the results are for a non-leveraged long-only portfolio of $1,000,000.
The algorithmic system analyzes past $N$ daily prices of given assets, constructs portfolio and observes cumulated performance for the next $N$ days. The portfolio simulation is based on the out-of-sample analysis. The simulation does not consider slippage, market impact and spreads but account for commissions, expense ratios and corporate events. We used Tiered commission plan of Interactive Brokers as of 1-Apr-2017 for simulation of both algorithms in MatLab 2015a package.

|  | Sharpe | Treynor x 0.01 | Return, % | Protection, % | STD, % | beta |
|---|---|---|---|---|---|---|
| Fractal biased (A) | 1.29 | 0.36 | 9.09 | 93 | 7.06 | 0.25 |
| Standard biased (B) | 1.19 | 0.32 | 8.91 | 92 | 7.49 | 0.28 |
| Naïve risk parity (C) | 0.9 | 0.24 | 8.18 | 86 | 9.4 | 0.36 |
| Benchmark (SPY) | 0.52 | 0.08 | 8.18 | 62 | 15.68 | 1 |
| Improvement A-B, % | 8 | 13 | 2 | 1 | 6 | 11 |

Table 3. Annual horizon, $1,000,000

|  | Sharpe | Treynor x 0.01 | Return, % | Protection, % | STD, % | beta |
|---|---|---|---|---|---|---|
| Fractal biased (A) | 1.31 | 0.45 | 9.68 | 91 | 5.10 | 0.15 |
| Standard biased (B) | 1.18 | 0.28 | 9.3 | 90 | 5.47 | 0.23 |
| Naïve risk parity (C) | 0.93 | 0.19 | 8.6 | 84 | 6.43 | 0.31 |
| Benchmark (SPY) | 0.5 | 0.06 | 8.26 | 64 | 11.36 | 1 |
| Improvement, % | 11 | 61 | 4 | 1 | 7 | 35 |

Table 4. Half-year horizon, $1,000,000

Capital protection is defined as the percentage difference between initial capital (i.e. 100%) and maximum drawdown of portfolio value in a single simulation. All metrics are in the annualized format. Relative percentage improvements are provided in the last raw. The decrease of the holding period strengthens the influence of inefficiency. Consequently, the advantage of fractal estimator of risk is more distinct. As a result, almost all metrics are improved despite more frequent rebalancing and higher execution costs. In this case, the system A beats the market on average by 1.4% for the twenty out-of-sample periods (see Table 4). The naive risk parity showed worst results for both horizons, which verified the advantage of proposed risk management scheme. Figure 1 represents absolute returns for the benchmark (SPY), standard biased risk parity, and fractal biased risk parity (please note that missing bars mean returns <1%). The most striking difference is noted for the 2007 half-year bar: -36% for SPY and -10% for system B in contrast to -6% for system A.

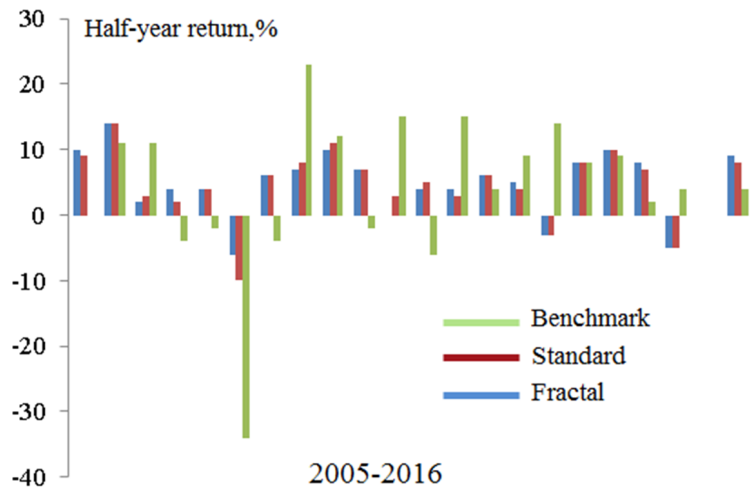

Figure 1. Absolute returns, half-year horizon

Below we represent the chart with reinvested returns for the half-year optimization and larger sample size of simulation. The bars correspond to the difference between the cumulated return of system A (Sh=1.31) and system B (Sh=1.18). The beginning of debt crisis (marked by black rectangular) is remarkable due to the significant growth of cumulated return.

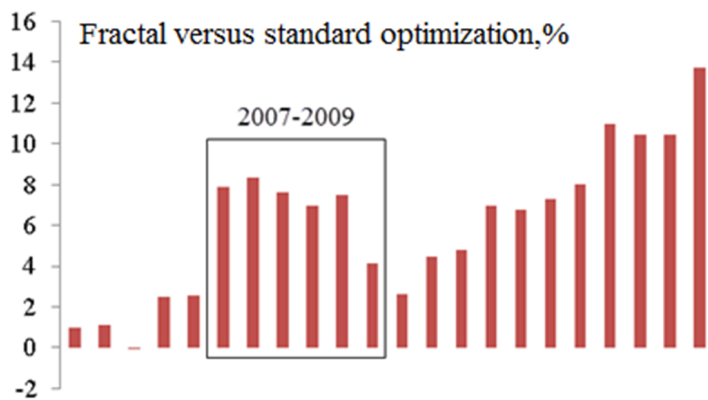

Figure 2. Difference A-B of cumulated returns, half-year horizon

However, the advantage of risk parity decreases as market efficiency recovers. Nonetheless, we observe the gradual increase in difference between systems A and B even in normal quasi-efficient market environment during the 10 years of observation. This advantage reaches 14% at the end of 2016. In spite of impressive results it should be noted that all performance metrics may be overestimated due to the limited sample size – we have only 20 returns for half-year optimization period. The further research of model applicability in 20-30 years historical database is recommended for accurate justification of model. For example, according to AQR research [6], the naive risk parity strategy based on stocks, bonds and commodities gives Sharpe ratio of 0.45 and annualized return of 11% over 39 years of historical test (January 1971 through December 2009). The Sharpe ratio is twice lower than the one simulated for naive risk parity in the current research. In Figure 3, the absolute cumulative return of fractal risk parity is compared to the SPY reinvestment strategy. It is remarkable that 2007-2008 advantage still plays the noteworthy role in the total performance of the strategy because it provides realistic estimation of tail risk for highly inefficient bearish environment. The provided results are contradictive to the MPT supporters who consider heavy tail events statistically insignificant for the long-term total performance of portfolio.

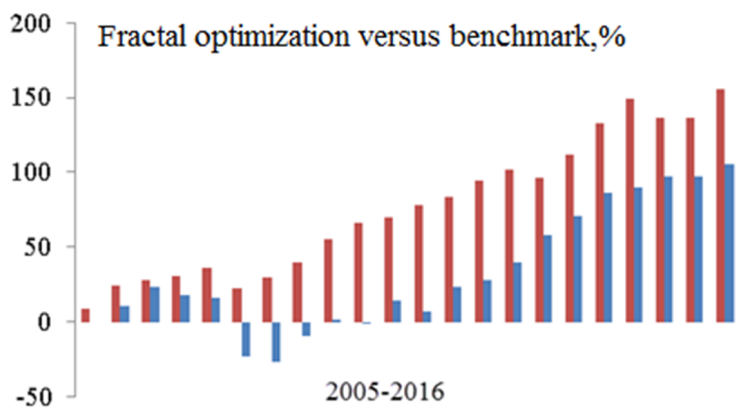

Figure 3. Absolute cumulated return, half-year horizon

**Conclusion**

In the current research, we investigated a fractal distribution of returns as a model for global asset allocation based on risk parity approach. We compared this approach to the traditional model of risk valuation, which implies normal distribution of returns and market efficiency hypothesis. The simulation results show that the proposed model improves portfolio metrics compared to the standard biased, naive risk parity and the benchmark. A portfolio with half-year optimization shows best results – its Sharpe ratio 2.6 times higher than that of the benchmark while average annual return is higher by 1.5%. The application of this approach in reinvestment cycle allows increasing an advantage of capital accumulation by 14% percent over the ten years. The improvement of performance metrics is most remarkable during the debt crisis of 2007-2008. It proves the advantage of tail risk models in herding market regimes when custom estimators of volatility are unstable. We suppose that the proposed approach may be valuable for managers that use original models of risk valuation. It may be incorporated in passive conservative investment strategies with wider universe of asset classes as well. We are positive that the model should be of great interest to the risk parity followers who tend to improve their strategies by more realistic view of abnormal market behavior.